# Single-scan scatter correction in CBCT by using projection correlation based view interpolation (PC-VI) and a stationary ring-shaped beam stop array (BSA)

Hao Yan, Xuanqin Mou, Yanbo Zhang, Maria Zankl

*Abstract*—In the scatter correction for x-ray Cone Beam (CB) CT, the single-scan scheme with moving Beam Stop Array (BSA) offers reliable scatter measurement with low dose, and by using Projection Correlation based View Interpolation (PC-VI), the primary fluence shaded by the moving BSA (during scatter measurement) could be recovered with high accuracy. However, the *moving* BSA may increase the mechanical burden in real applications. For better practicability, in this paper we proposed a PC-VI based single-scan scheme with a ring-shaped stationary BSA, which serves as a *virtual moving* BSA during CB scan, so the shaded primary fluence by this stationary BSA can be also well recovered by PC-VI. The principle in designing the whole system is deduced and evaluated. The proposed scheme greatly enhances the practicability of the single-scan scatter correction scheme.

## I. INTRODUCTION

In CBCT imaging, there are generally two types of scatter correction strategies: model-based and measurement-based. A typical measurement-based method is the beam stop array (BSA) method [1, 2]. It is reliable with respect to the scatter measurement, but the (primary) beam stop based measurement causes that a part of the primary fluence is lost. Hence, in total two sets of scans are needed to get complete projections. A relatively dose-practical BSA method is that of Ning *et al* (2004) [2], where a sparse-view scan with BSA is performed preceding a full-view normal scan (without BSA). The scatter fluence in the sparse views is estimated by spatial interpolation based on the measurements. The scatter fluence in other views is estimated by angular interpolation of the sparse-view scatter fluence estimates. Thus, in this method there exists a compromise between dose consideration and the scatter estimation accuracy in other views, i.e. accuracy is decreasing with sparser views; but to estimate more accurately by increasing the views, dose would be increased.

Giving a high priority to the dose consideration, the single scan scheme with moving BSA was developed [3]. In this method, the scatter measurement is performed in each view, so the scatter estimation is quite accurate; however, the adopted spatial interpolation (SI) performs not quite well in restoring the BSA-shaded primary fluence, although the BSA was designed as moving in a 2D raster mode, in order to decrease the cumulated SI error. Recently, the moving BSA method has attracted many researchers [4~6], and a theory breakthrough has been reported in restoring the BSA-shaded primary fluence, i.e. the projection correlation based view interpolation (PC-VI) [6]. PC-VI far outperforms the traditional SI, and it works well when the BSA shadows between neighboring views do not overlap (PC-VI requirement). Consequently, under the PC-VI framework, the moving BSA is more aimed at fulfilling PC-VI requirement, rather than at reducing the cumulated SI error, and hence, great flexibility in the BSA moving mode became possible, e.g. a 1D moving and a rotating mode (Fig. 1). The PC-VI based moving BSA method has achieved a balance on the dose practicability, accuracy in both scatter measurement and primary restoration, and flexibility of BSA movement.

Even so, the *moving* BSA still means a heavy demand for the mechanical implementations. In this paper, we present the design of a stationary ring-shaped BSA that replaces the moving BSA. We deduce and investigate the principle for the ring-shaped BSA design in the context of PC-VI. With this ring-shaped BSA, an effect equivalent to using a moving BSA is generated, so the PC-VI requirement is fulfilled and PC-VI works well in restoring the BSA-shaded primary fluence. The proposed method is joining the advantages of both the BSA and the moving BSA methods reviewed above, namely, the BSA is not necessarily moving; and a single-scan is enough for both the

This work was supported in part by the National Natural Science Foundation of China (NSFC) under Grant No. 60551003 and 60472004, and the Doctoral Fund of the Ministry of Education of China under Grant No. 20060698040.

Hao Yan, Xuanqin Mou (corresponding author) and Yanbo Zhang are with the Institute of Image Processing & Pattern Recognition, Xi!an Jiaotong University, 710049, Xi!an, P.R.China (e-mail: yhhere@126.com, xqmou@mail.xjtu.edu.cn, yanbozhang007@163.com).

Maria Zankl is with the Helmholtz Zentrum München " German Research Center for Environmental Health, Department of Medical Radiation Physics & Diagnostics, Neuherberg, Germany (e-mail: zankl@helmholtz-muenchen.de).

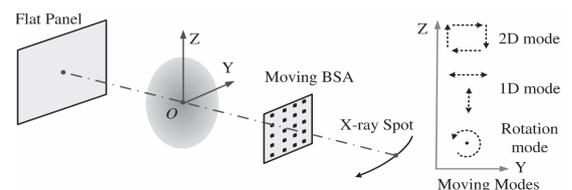

Fig. 1 The PC-VI based moving BSA setup

scatter measurement and the primary fluence acquisition with high accuracy. For this reason, the proposed method is supposed to be a large improvement to the BSA measurement based scatter correction.

## II. Principle for the Stationary Ring-Shaped BSA Design

### A. Seminal work on PC-VI

Denoting the x-ray focal spot as $\vec{r}$ and the detector cell as $\vec{u}$, the weighted x-ray transform $g(\vec{r}, \vec{u})$ of an object $f$ is:

$$g(\vec{r}, \vec{u}) = \int_0^1 f(\vec{r} + t(\vec{u} - \vec{r})) dt = X(f(\vec{r}, \vec{u}))/|\vec{u} - \vec{r}|. \quad (1)$$

where $X(f(\vec{r}, \vec{u}))$ is the CT data, the normalized line integral of $f$. Under a flat panel CBCT setup, $\vec{r} = (r, \beta, \dot{z})$, $d$, and $\vec{u} = (u, v, z)$, where $u$ and $v$ are the 2D detector coordinates; $\beta$ is the azimuth angle; $r$ and $d$ are the source-to-centre and centre-to-detector distances; $z$ is the longitudinal coordinate. In the following text, $g(\vec{r}, \dot{r}, d)(u, v, z)$ is abbreviated as $g(\beta, \dot{z}, u, v)$, or further as $g$. Denoting $g_{xy}$ as the partial differential of $g$ to variables $x$ and $y$, and by using $^*$ to refer to the Fourier transform, and simplifying $g^*(\beta, k_1, k_2)$ as $g^*$, the essential formula for PC-VI is:

$$g^*_\beta - \frac{j}{r+d}\left[\left(2\frac{k_1}{k_2}g^*_{k_1} - k_2 g^*_{k_1 k_2}\right) + k_1 g^*_{k_2 k_2}\right] k_2 = 0. \quad (2)$$

It indicates that the high spatial frequency information of $g$ is mainly contained in neighboring angular projections ($g_\beta$), besides in $g$ itself. Based on (2), the PC-VI was developed [6]. The PC-VI requirement is quoted as below:

*PC-VI requirement: PC-VI works when angularly nearby pixels are complete, e.g. the calculation of $g(\beta + d\beta, \dot{u}, v)$ requires the knowledge of $g(\beta, u, v)$ or $g(\beta + d\beta, \dot{u}, v)$, where $\dot{u}$ represents pixel u and its neighborhood.*

According to the PC-VI requirement, the locations of shaded pixels are supposed to be changing between neighboring views; hence a moving BSA is required instead of a stationary one, since which always blocks pixels at the same position.

### B. Ring-shaped BSA design: the basic idea

We notice that a traditional BSA has a planar shape and is placed before the x-ray emitter; thus, it rotates together with x-ray source [2] [3]. As a result, when this BSA is static, it blocks the same positions of the flat panel in each view. According to this observation, we have designed a stationary BSA that does not rotate with the x-ray source. The design sketch of this idea is shown in Fig. 2.

The stationary BSA is a thin-walled PMMA ring-shell in which embedded with iso-tropically distributed lead balls. As shown in Fig. 3(a), this ring-shaped BSA can be placed fixedly between the trajectories of the x-ray source and the flat panel when $|OH| < l < r$, i.e. $(d^2 + h_U^2)^{1/2} < l < r$. Different from the angularly rotating planar BSA, relative displacements between the blocker (and shaded pixels) positions in neighboring views are generated by this ring-shaped design, and this is coincident with the PC-VI requirement. To satisfy the PC-VI requirement perfectly, a deduction for the parameters tuning is necessary, and the related symbols are given in the caption of Fig. 3.

Fig. 2 Design sketch of the ring shaped BSA setup

### C. Ring-shaped BSA design: parameters deduction

We make deductions with the basic ideas as illustrated in Fig. 3(b) and (c). For an arbitrary lead ball centered at $B(l\cos\psi, l\sin\psi, \zeta)$, we need to locate its shaded pixels in two adjacent views ($\xi_0, \xi_1$). This involves three steps: (1) Calculate the position of $\xi_0$ and $\xi_1$ under the global coordinates ($XYOZ$); (2) transform the results of (1) to the local coordinates ($UO'V$) to get the local position of the shade pixels; and (3) calculate the amplification factor $a(\beta)$ to get the shadow size, and then get the shadow boundary e.g. $\eta_{0-}$, $\eta_{0+}$, $\eta_0^Z$ and $\eta_0^!$ for shadows centered at $\xi_0$ (and similarly $\eta_{1-}$, $\eta_{1+}$, $\eta_1^Z$, $\eta_1^!$ for shadows centered at $\xi_1$), as seen from Fig. 3(c).

On this basis, we can study the condition under which these two shadows do not overlap, to satisfy the PC-VI requirement.

For arbitrary $\vec{r} = (r\cos\beta, r\sin\beta, z_0 + p\beta/2\pi)$, the line $lB$ is:

$$\frac{x - l\cos\psi}{r\cos\beta - l\cos\psi} = \frac{y - l\sin\psi}{r\sin\beta - l\sin\psi} = \frac{z - \zeta}{z_0 - \zeta + p\beta/2\pi - \zeta}. \quad (3)$$

The related flat panel plane $UO'V$ is:

$$\cos\beta[x - d\cos(\beta + \pi)] - \sin\beta[y - d\sin(\beta + \pi)] = 0. \quad (4)$$

Denoting the angle between $UO'V$ and $lB$ as $\lambda$

$$\lambda = \frac{d}{2} - \langle n_{UO'V}, \quad (5)$$

(a)

(b) $B(l\cos\psi, l\sin\psi, \zeta)$   (c)

Fig. 3 Parameters illustration (a) Left: 3D view; right: top view. The radius of the ring-shell and each lead ball are equal to $l$ and $\sigma$ respectively. The azimuth angle of the lead ball is denoted as $\psi$. The view interval is denoted as $d\beta$. The pitch of the spiral source trajectory is denoted as $p$. The half detector length in $U$ and $V$ direction is denoted as $h_U$ and $h_V$. (b) The deduction illustration. $a$ represents the amplification factor. $B$ is the lead ball centre, and the azimuth interval between the adjacent lead balls is $\psi_m$. (c) Local details of (b). The angle between $\vec{r}$ and the flat panel is denoted by $\lambda$. $\eta_{0-}$ and $\eta_{0+}$ represent the shadow boundary in $U$ direction; $\eta_0^Z$ and $\eta_0^!$ represent the shadow boundary in $V$ direction.

where $\mathbf{l}_B$ is the direction vector of $\overline{AB}$; and $\mathbf{n}_{UO'V}$ is the normal vector of $UO'V$,

$$\mathbf{l}_B = (l\cos\alpha - l\cos\beta, -\sin\alpha - l\sin\beta, z_0 + p/2\pi \cdot \beta - r) \quad (6)$$

$$\mathbf{n}_{UO'V} = (\cos\beta, \sin\beta, 0). \quad (7)$$

From (5) ~ (7), we get

$$\sin\gamma = \frac{|r - l\cos(\alpha-\beta)|}{|\mathbf{l}_B|}. \quad (8)$$

As shown in Fig. 3(c), the magnification factor to determine the size of the lead ball shadow can be approximated by:

$$a \approx \frac{|\mathbf{l}_A|/|\mathbf{l}_B|}{\sin\gamma} = \frac{|\mathbf{l}_A|}{|r - l\cos(\alpha-\beta)|}. \quad (9)$$

If one transforms the global coordinate $XYOZ$ to the local coordinate $UO'V$, (3) can be expressed under $UO'V$. Then we can get the local coordinate $(u, v)$ of arbitrary $\mathbf{l}_A$:

$$u = \frac{l(\beta - d)\sin(\alpha-\beta)}{r - l\cos(\alpha-\beta)}, \quad (10)$$

$$v = z_0 - \frac{(l_0 - d)p/2\pi \cdot \beta - r \cdot (l_0 - d)}{(r - l\cos(\alpha-\beta))}. \quad (11)$$

By transforming (10) and (11) back to $XYOZ$ and substituting them into (9) with the global coordinates of $A$ and $B$, we get:

$$a = \frac{(\beta - d)\sqrt{r^2 - 2rl\cos(\alpha-\beta) + l^2 + (l_0 \cdot p/2\pi \cdot \beta - r)^2}}{|r - l\cos(\alpha-\beta)|^2}. \quad (12)$$

Since the lead balls are distributed iso-tropically on the ring shell and $B$ is selected arbitrarily, we assume $z_0=0$ and select two specific x-ray focus position $\beta=0$ and $\beta=d\beta$, to simplify the deduction. By substituting $\beta=0$ and $\beta=d\beta$ into (10), (11) and (12), respectively, we get the local coordinates $u$, $v$ and the magnification factor $a$ for the line $l_s$-$l_s$ and $l_s$-$l_s$. On this basis, we can estimate the shadow boundary, e.g. $u(l_s)$, $u(l_s) \pm a(l_s) \cdot g$. Recalling Fig.3 (b) and (c), either $u(l_s) < u(l_s)$ or $v(l_s) < v(l_s)$ should be correct to fulfill PC-VI requirement, and note that the later does not hold unless pitch ($p$) is very large, so we make a further deduction for $u(l_s) < u(l_s)$:

$$u(l_0) + u(l_1) + 2g \cdot (l_0) + (l_0 - a(l_1)) \cdot \text{"} \quad (13)$$

By substituting $u$, $v$ and $a$ of the line $l_s$-$l_s$ and $l_s$-$l_s$ into (13), we get the principle for this ring-shaped BSA design:

$$G(\alpha, d, d\beta, h_U, h_V, l, g, m, n) =$$

$$\frac{1}{2}l\left(\frac{\sin\alpha}{-l\cos\alpha} - \frac{\sin(\alpha-d\beta)}{r-l\cos(\alpha-d\beta)}\right) \quad (14)$$

$$+g\cdot\left(\frac{\sqrt{r^2-2rl\cos\alpha+l^2+n^2}}{|r-l\cos\alpha|^2} + \frac{\sqrt{r^2-2rl\cos(\alpha-d\beta)+l^2+(p/2\pi\cdot d\beta-r)^2}}{|r-l\cos(\alpha-d\beta)|^2}\right) > 0$$

### D. *Principle for the ring-shaped BSA design: system setup*

For the ring-shaped BSA design, both $(d^2+h_U^2)^{1/2}<l<r$ and (14) should be satisfied. To make the design practicable, it is

TABLE I PARAMETERS TUNING

| Parameters | Values |
|---|---|
| Fixed | $h_U=345$, $h_V=100$, $g=1\%$ |
| Varying | $n = 0.0920*(750-l)$, $\alpha \in [-n, n]/4 : n_s]$ |
| | $m = 0.1636$, $m \in [-m, m]/15 : m_s]$ |
| | $l \in [485:5:520]$ $d \in [300:5:400]$ |
| | $r \in [650:5:950]$; $d\beta \in [1:3:20]$, $n \in [540:9:1080]$ |
| Temporarily fixed | $r=750$, $d=337.5$, $l=485$, $d\beta=180$, $m=m_s$, $n=n_s$ |

(The unit for $d\beta$ and $m$ is deg, unit for other parameters is *mm*)

necessary to study on the parameters tuning.

Firstly, a general investigation is performed by fixing basic parameters (row 1, table I) and tuning other parameters (row 2, table I). Each time only one parameter is varying and others are temporarily fixed (row 3, table I). By this procedure, two values are investigated, the first is G in (14); the second is the lead ball shadow size (denoted as S, and $S \approx 2a \cdot g$). The former is for fulfilling the PC-VI requirement, and the later is for better PC-VI performance, considering that restoration is much easier when a smaller fraction of detector is blocked. From the general investigation, we would like to report the results directly: the key parameters influencing G are $d\beta$ and $r$; and (14) is more likely to be fulfilled with smaller $r$ and larger $d\beta$; we also learn that the key parameters influencing S is $r$ and $l$, and fewer pixels are shaded with larger $r$ and smaller $l$.

Secondly, a detailed investigation of the key parameters is performed by varying $(d\beta, r)$ and $(r, l)$, with other parameters fixed (row 3, table I). As shown in Fig. 4(a), when G=0, $d\beta$ is equal to around $2\pi/930$, nearly invariable for varying values of $r$, i.e. G is mainly determined by $d\beta$, and the critical point of total view number is ~930; As shown in Fig. 4(b), both $r$ and $l$ determine S, and tuning $r$ seems more effective due to the limited range of $l$, recalling $(d^2+h_U^2)^{1/2}<l<r$.

### E. *CBCT setups with embedded exemplary ring-shaped BSA*

The exemplary ring-shaped BSA contains 400×8 lead balls: the 400 columns are equally distributed over 360 degrees; and the 8 rows are quasi-equally distributed along the axis direction. Its projection image on a large enough flat panel is shown in Fig. 5(a), where the physical size of the flat panel is marked by the vertical lines. Fig. 5(b) displays two adjacent projections and the result when they are super-posited. Based on the former studies, we assign $r=750$, $d=337.5$, $l=485$; two exemplary setups with $d\beta$ of (Ⅰ) $d\beta=180$ and (Ⅱ) $d\beta=1080$, are adopted. Accordingly, the top-left quarters of the super-posited views are shown in Fig. 5 (c), where the lead ball shadows in adjacent views do not overlap for (Ⅰ), and slightly overlapped for (Ⅱ).

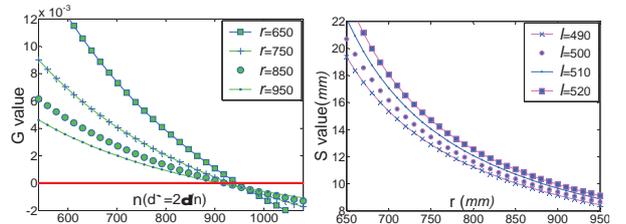

Fig. 4 Detailed investigations: (a) G with $(d\beta, r)$ (b) S with $(r, l)$.

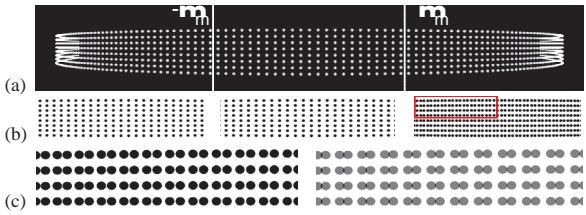

Fig. 5 Exemplary setups (a) Full view image of the ring shaped BSA on a large flat panel. (b) Adjacent projections and the super-posited results. (c) The top-left quarter of the super-posited projections under the setup of ( ), ( ).

### F. Evaluations and results

Circular CBCT scans of the *Visible Human* [7, 8] have been simulated with the hybrid technique in [6]. Three setups are investigated, i.e. ( ) , ( ) and a laterally moving BSA [6] as a reference. With the BSA measurements, the scatter corrections have been performed, as in [3]. The accuracy is nearly identical for the three setups because in which lead balls are distributed closely enough and rather similarly (~10% pixels are shaded).

On this basis, PC-VI is employed to restore the shaded primary information, and its performance can be evaluated independently in the projection domain with

$$|Error|_{PC-VI} = mean_\lambda \left\{ \frac{\sum_{u,v \in \Omega} |View_{PC-VI}(\lambda,u,v) - View_0(\lambda,u,v)|}{\sum_{u,v \in \Omega} |View_0(\lambda,u,v)|} \right\} \quad (15)$$

where $\Omega$ represents the set of the shaded pixels. Similarly with (15), $|Error|_{SI}$ is also calculated for using SI. From the results in table . It shows that with the stationary ring-shaped BSA, the performance of PC-VI is similar to the setup with the moving BSA. The error in setup ( ) is slightly larger than that of ( ).

Further investigations have been performed in the reconstructed images of a representative slice with severe blocking (passing through most of the shadow centers). As demonstrated in Fig. 6, PC-VI works well for both setups and the results are generally similar to the blocking-free images.

### III. DISCUSSION AND CONCLUSION

A formula ((14)) based principle is proposed for designing a stationary ring-shaped BSA to replace the physical moving BSA. A study on parameters tuning is performed (Fig. 4), and then a guide in designing the ring-shaped BSA is formulated. For the exemplary setups (Fig. 5), PC-VI works well, even for setup ( ) where the shadows slightly overlap (Fig. 6). Compared with a moving BSA, the accuracy of using a ring-shaped BSA is generally the same (table ). Thus, using a ring-shaped BSA instead of a moving BSA is indeed feasible.

As observed from table , PC-VI performs better under setup ( ) than ( ), although the later is provided a denser view sampling (and under which PC-VI performs better, see [6]). The reason might be, compared with setup ( ), the case of ( ) is not so perfect that PC-VI requirement is violated slightly; Thus PC-VI is less powerful. On the other side, no evident degradation is observed in the images from Fig. 6. The reason might be the view sampling under setup ( ) is finer, which alleviates the image distortions and counterbalances the negative effect of violating PC-VI requirements, since in the case of ( ), the total view number is slightly increased over the critical point. And that also explains why fewer steaks have been observed under setup ( ) for the SI case.

Although (14) is derived for the flat panel based setup, it is straight-forward for the so-called *native geometry* with a curved detector, and under which the design is much easier, since the radius range of the ring is much larger, i.e. $d<l<r$. In our example, blockings are concentrated in only a few slices, since it was our aim to maximize the primary blocking errors in typical slices, to evaluate our method under extreme conditions. The blockings can be adopted as other distributions, and under which when the blockings are averaged over more slices, better results can be achieved. Furthermore, our results are also applicable for beam stop strips/lines.

In conclusion, the stationary ring-shaped BSA serves like a virtual moving BSA in the CB scan, so it performs well as a replacement of the previously used moving BSA, establishing a PC-VI based single-scan scheme with more practicability.

TABLE   COMPARISONS: STATIONARY BSA AND MOVING BSA

| appurtenance | setup | Mean absolute error | |
|---|---|---|---|
| | | $|Error|_{SI}$ | $|Error|_{PC-VI}$ |
| stationary BSA | ( ) | 1.279% | 0.346% |
| | ( ) | 1.276% | 0.416% |
| moving BSA | Lateral mode[6] | 1.180% | 0.399% |

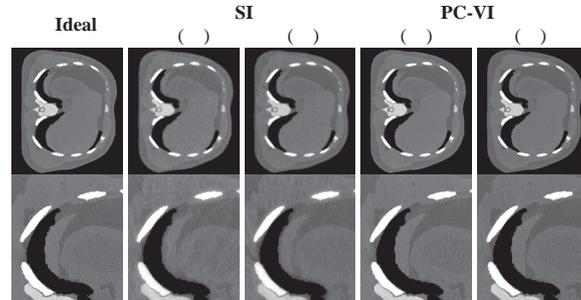

Fig. 6 Representative slice: the reconstructed images (1st row) and the enlarged details (2nd row). Displayed window ([-500, 900] HU)